\documentclass{JHEP}
\usepackage{graphicx}
\usepackage{epsfig}
\usepackage{amssymb}
\usepackage{amsmath}
\usepackage{float}

\newcommand{\bea}{\begin{eqnarray}}
\newcommand{\eea}{\end{eqnarray}}
\newcommand{\bean}{\begin{eqnarray*}}
\newcommand{\eean}{\end{eqnarray*}}
\newcommand{\nn}{\nonumber \\}

\def\W #1{\widetilde{#1}}
\def\WH #1{\widehat{#1}}

\def\eref#1{(\ref{#1})}

\def\a{{\alpha}}

\def\b{{\beta}}

\def\Sl{\sum\limits}
\def\Label#1{\label{#1}%
  \smash{\hbox to0pt{\raise1ex\hbox{\tiny[#1]}\hss}}}

\allowdisplaybreaks
\title{Note on Cyclic Sum and Combination Sum of Color-ordered Gluon Amplitudes }

\author{Yi-Jian Du${}^{a}$, Bo Feng${}^{a,b}$, Chih-Hao Fu${}^{b}$    \\
$^a$\small Zhejiang Institute of Modern Physics, Zhejiang
University, Hangzhou, 310027, P. R. China\\$^b$\small Center of
Mathematical Sciences, Zhejiang University, Hangzhou, 310027, P. R.  China \\
}

\date{\today}
\abstract{Continuing our previous study \cite{Du:2011se} of
permutation sum of color ordered  tree amplitudes of gluons, in this
note, we prove the large-$z$ behavior of their cyclic sum and the
combination of cyclic and permutation sums under BCFW deformation.
Unlike the permutation sum, the study of cyclic sum and the
combination of cyclic and permutation sums is much more difficult.
By using the generalized Bern-Carrasco-Johansson (BCJ) relation, we
have proved the boundary behavior of cyclic sum with nonadjacent
BCFW deformation. The proof of
 cyclic sum with adjacent BCFW deformation is a little bit simpler,
where only Kleiss-Kuijf (KK) relations are needed. Finally we have
presented a new observation for partial-ordered permutation sum and
applied it to prove the boundary behavior of combination sum with
cyclic and permutation.
}

\keywords{Gauge Symmetry
, Duality in Gauge Field Theories}

\begin{document}

\section{Introduction}

On-shell recursion relation  for tree-level gluon amplitudes
\cite{Britto:2004ap, Britto:2005fq} has been shown to be important
not only for real calculations, but also for theoretical
understanding of many important properties such as the BCJ relaton
\cite{Bern:2008qj}\footnote{The BCJ relation has first been proved
in string theory \cite{BjerrumBohr:2009rd,Stieberger:2009hq}, and
then in field theory \cite{Feng:2010my, Chen:2011jxa}.} and
Kawai-Lewellen-Tye relation
\cite{Kawai:1985xq,Bern:1998sv,BjerrumBohr:2010ta,
BjerrumBohr:2010zb, BjerrumBohr:2010yc}\footnote{See also recent
review \cite{Sondergaard:2011iv}.}. To establish the on-shell
recursion relation, understanding of large-$z$ behavior (or  the
``boundary behavior'')
 of amplitude under
BCFW-deformation on a chosen pair of particles $(i,j)$
\bea  p_i\to p_i-zq, ~~~p_j\to p_j+z q,~~~q^2=q\cdot p_i=q\cdot
p_j=0~~~~\label{BCFW-def}\eea
becomes crucial. However estimating boundary behavior is not so easy
and a naive analysis from Feynman diagrams could often lead to wrong
conclusions. A careful  analysis was done by Arkani-hamed and
Kaplan in \cite{ArkaniHamed:2008yf}, where because  $zq\to \infty$,
the whole amplitude can be considered as scattering of particles $i,j$
from soft background constructed by other particles. For practical
purposes it is  considerably simpler if 
$A(z)\to 0$ as 
 $z\to \infty$,
so that the boundary contribution is simply zero. Furthermore, if the amplitude
has even better asymptotic behavior $A(z)\sim {1\over z^k},~k\geq 2$, it is possible
to derive more relations  in addition to the standard on-shell
recursion relation. These ``bonus'' relations were discussed in
\cite{Benincasa:2007qj, ArkaniHamed:2008yf,
Spradlin:2008bu,Badger:2008rn, Feng:2010my,Badger:2010eq,
He:2010ab}, where their usefulness was demonstrated from various
aspects.

Because of its importance, it is desirable to have better understanding
to this problem. Recently, the boundary behavior of gluon amplitude
under deformation (\ref{BCFW-def}) has been carefully studied  by
Boels and Isermann in \cite{Boels:2011tp, Boels:2011mn}, where some
new behaviors were observed.  Among them, the following two
statements are particularly intriguing:
\bea \sum_{perm~\a} A_n (i, \{\a\}, j, \{\b\})\to
\xi_{i\mu}(z)\xi_{j\nu}(z){ G^{\mu\nu}(z)\over z^{k}},~~~k=\left\{
\begin{array}{ll} n_\a, &  i,j~not~nearby \\ n_\a-1,~~~ &  i,j~~nearby
\end{array} \right.~~~\label{Boels-1}\eea
and\footnote{The second statement requires the number in the set
$\a$ to be equal or larger than two.}~\footnote{As mentioned by
Boels and Isermann in \cite{Boels:2011mn}, though the behavior of
permutation sum and combination sum could be  checked up to and
including $z^{-2}$, the general proofs of these behaviors were hard
to given in this way. We should also notice that the adjacent cyclic
sum is considered as a special case of combination sum in
\cite{Boels:2011mn}. However, the non-adjacent cyclic sum cannot be
regarded as a special case of the combination sum and this case was
not checked in \cite{Boels:2011mn}.}

\bea \sum_{cyclic~\a} A_n (i, \{\a\}, j, \{\b\})\to
\xi_{i\mu}(z)\xi_{j\nu}(z){ G^{\mu\nu}(z)\over z^{k}},~~~k=\left\{
\begin{array}{ll} 2, &  i,j~not~nearby \\ 1,~~~ &  i,j~~nearby
\end{array} \right.~~~\label{Boels-2}\eea
where $n_\a$ is the number of elements in set $\a$, $\xi$ is the
polarization vector and $G_{\mu \nu}$ is given by
\cite{ArkaniHamed:2008yf}
\bea G_{\mu \nu}= z\eta^{\mu\nu} f(1/z)+ B^{\mu\nu}(1/z)+{\cal
O}(1/z)~.\eea
These two observations can combine together and following sum has
the large-$z$ behavior
\bea \Sl_{\sigma\in Z(\{2,...,i-1\})}\Sl_{\rho\in
P(\{i+1,...,n\})}A(\WH1,\sigma,\WH i,\rho) \to
\xi_{1\mu}(z)\xi_{i\nu}(z){ G^{\mu\nu}(z)\over
z^{n-i+1}}~,~~\label{Boels-com}\eea
i.e., it is one ${1\over z}$ suppressing comparing to the pure
permutation sum of \eqref{Boels-1}.

The above results by Boels and Isermann were observed from explicit
analysis of Feynman diagrams. Despite intuitive and conceptually
straightforward, such approach requires order by order cancelations,
which is unfortunately difficult to show and the fundamental
mechanism behind these cancelations and the
 better convergent
behavior is far from transparent. Furthermore, as we have mentioned,
better divergent behavior could often imply extra relations among
amplitudes, such as BCJ and KLT relations, it is crucial and natural
 to ask {\it do these two new statements lead to some un-discovered
nontrivial relations among color-ordered tree amplitudes}?

To better understand the reasons behind cancelations found by Feynman
diagram analysis and answer the question raised in previous
paragraph, we study boundary behavior of color-ordered amplitudes
from another point of view. In \cite{Du:2011se} we have studied the
first statement (\ref{Boels-1}) and showed that by using
Kleiss-Kuijf (KK) relation and fundamental BCJ relation, statement
(\ref{Boels-1}) can be derived easily. In this note we continue our
study on the second statements (\ref{Boels-2}), \eqref{Boels-com}.
Surprisingly we find that comparing
 to the first statement, the second statement is technically
 much more difficult to investigate. Besides the familiar Kleiss-Kuijf (KK)
relation and fundamental BCJ relation, an extensive application of
generalized BCJ relation\cite{Chen:2011jxa} is required\footnote{ KK
relation and generalized BCJ relation in fact can be generated by
only using fundamental BCJ relation in addition with cyclic
symmetry\cite{Ma:2011um}. }. This is because fewer symmetries
are possessed by the cyclic sum as opposed to those by
the permutation sum, and therefore the greater
technical challenge is involved. To be
able to understand these new nontrivial technical points, before
each general proof we provide an example to demonstrate in
details.

Our results make the following statements more transparent: First, the
observed cancelations in Feynman diagrams after the cyclic or
permutation sum are natural consequences of the well known boundary
behavior under the  BCFW-deformation. Secondly, since the
bonus relations of a pair deformation, i.e., the BCJ relations and
their generalizations, have been found, there are no new nontrivial
bonus relations implied by the better boundary behavior of cyclic or
permutation sum.

We would like to emphasize that although there are no new bonus relations for
cyclic or permutation sum, {\it there do exist many nontrivial
applications for cyclic or permutation sum}. Some nontrivial
examples have been given in \cite{Boels:2011tp, Boels:2011mn}, where
vanishing of box, triangle or bubble coefficients of one-loop
amplitudes has been understood from this point of view. Thus it is
desirable to study boundary behavior for other cyclic and
permutation combinations. A new result by our method
 is that we have
observed the large-$z$ behavior of another type of sum, i.e., the partial-ordered
permutation sum
\bea \Sl_{\sigma\in
P(O\{2,...,l\}\bigcup\{l+1,...,i-1\})}A(\WH1,\sigma,\WH i,...,n)\sim
\frac{1}{z}\Sl_{\sigma'\in P(\{l+1,...,i-1\})}A(\WH1,\sigma',\WH
i,...,n)~~~\label{by-product}\eea
where the sum is over all permutations of elements $\{2,3,...,i-1\}$
under one condition: the relative ordering in the subset
$\{2,...,l\}$ is kept. An application of our new result
\eref{by-product} is to prove \eref{Boels-com} given by Boels and
Isermann. Other possible applications of \eref{by-product} could be on
simplifying coefficients of loop amplitudes. Also since our proof
uses only KK and (generalized) BCJ relations, which are also true
for ${\cal N}=4$ SYM theory, all results in this note are
automatically true for ${\cal N}=4$ SYM theory. Finally, it is
natural to generalize our method to other situations, such as to string
theory \cite{string} or Witten's diagram \cite{Witten-diagram} where
BCFW on-shell recursion relation has been applied.

The plan of this note is the following: In section two, we prove
the large-$z$ behavior for cyclic sum under  non-adjacent BCFW
deformation while in
section three we prove the large-$z$ behavior for cyclic sum under adjacent
 BCFW deformation. We present the proof for a new observation concerning
 the partial-ordered permutation
 sum \eqref{by-product} in section four
and finally using the result in section five
 we prove the combination sum given by \eqref{Boels-com}.

\subsection{Some backgrounds}

For self-completeness we review the formulas needed in our proofs. The
first one is the  Kleiss-Kuijf (KK) relation, which was first
conjectured in \cite{Kleiss:1988ne} and later proved in
\cite{DelDuca:1999rs}. The formula reads
\bea  A_n(1,\{\a\}, n,\{\b\}) = (-1)^{n_\b}\sum_{\sigma\in
P(O\{\a\}\bigcup O\{\b^T\})} A_n(1,\sigma,
n)~,~~~~\label{KK-rel}\eea
where the $P(O\{\a\}\bigcup O\{\b^T\})$ sum is to be taken over all
permutations of set $\a \bigcup \b^T$ whereas the relative ordering
in  sets $\a$ and $\b^T$ ( which is the reversed orderings of set
$\b$) are preserved. The $n_\b$ here is the number of elements in
set $\b$ . One non-trivial example with six gluons is given as the
following
\bea A(1,\{ 2,3\},6,\{4,5\}) & = & A(1,2,3,5,4,6)+ A(1,2,5,3,4,6)+
A(1,2,5,4,3,6)\nn & & +
A(1,5,4,2,3,6)+A(1,5,2,4,3,6)+A(1,5,2,3,4,6)~.
~~~\label{KK-6-point}\eea

The second formula we need is  the generalized BCJ relation given by
\cite{Chen:2011jxa}
 \bea \Sl_{\{\sigma\}\in P(O\{\alpha\}\bigcup
O\{\beta\})}\Sl_{i=1}^{n_\b}
\Sl_{\xi_{\sigma(J)}<\xi_{\sigma(\beta_i)}}s_{\beta_iJ}A_n(1,\{\sigma\},n)=0,~~~\label{gen-BCJ}
 \eea
where $(n-2)$'s elements have been divided into two subsets $\a, \b$
arbitrarily.
In the sum, the
position of an element $t$ in a given ordering $\sigma$ is denoted
by $\xi_{\sigma(t)}$ with the convention that the position of particle $1$
is defined as $\xi_{\sigma(1)}=0$, thus the sum
$\Sl_{\xi_{\sigma(J)}<\xi_{\sigma(\beta_i)}}s_{\beta_iJ}$ is over all elements
at the left hand side of $i$-th element in the set $\b$. One example
with six gluons is given as following
\bea 0 & = & [(
s_{41}+s_{42}+s_{43})+(s_{51}+s_{52}+s_{53}+s_{54})]A(1,
2,3,4,5,6)\nn & & + [( s_{41}+s_{42})+(s_{51}+s_{52}+s_{53}+s_{54})]
A(1,2,4,3,5,6)\nn & & + [(
s_{41}+s_{42})+(s_{51}+s_{52}+s_{54})]A(1,2,4,5,3,6)\nn & & +[(
s_{41})+(s_{51}+s_{52}+s_{53}+s_{54})] A(1,4,2,3,5,6)\nn & & +[(
s_{41})+(s_{51}+s_{52}+s_{54})]A(1,4,2,5,3,6)+ [(
s_{41})+(s_{51}+s_{54})]A(1,4,5,2,3,6)\eea
where the set $\a=\{2,3\}$ and the set $\b=\{4,5\}$. The generalized
BCJ relations will be used extensively in our current paper.

\section{The non-adjacent case of cyclic sum}

In this section we  prove the large-$z$ behavior of cyclic sum
for non-adjacent case, i.e.,
\bea \sum_{cyclic~(2,...,i-1)} A_n (\WH 1, \{2,..,i-1\}, \WH i,
i+1,...,n)\to \xi_{1\mu}(z)\xi_{i\nu}(z){ G^{\mu\nu}(z)\over
z^{2}}~~\label{Not-adj}\eea
with $4\leq i\leq n-1$. For the case $i=4$, the cyclic sum of two
elements is same as the permutation sum of them and the large-$z$
behavior can be read off from (\ref{Boels-1}) directly, which is
exactly (\ref{Not-adj}) and has been proved in \cite{Du:2011se}.
Having established that (\ref{Not-adj}) is true for $i=4$, we will
try to prove general $i$ by induction.

Comparing to the proof given in \cite{Du:2011se}, the behavior of
cyclic sum is not as good as the behavior of permutation sum. Thus
the method in \cite{Du:2011se} can not be applied directly. New
inputs must be cooperated, such as generalized BCJ relations and how
to split the cyclic sum into different types according to their
boundary behaviors. Because these complexities, before giving the
general proof, we will use the example $i=5$ to demonstrate our
idea.

\subsection{The example with $i=5$}

The first step is to use the generalized BCJ relation to rewrite
each term in the summation (\ref{Not-adj}). For example, for $A(\WH
1, 2,3,4,\WH 5, 6,...,n)$, if we choose the set $\b=\{2,3,4\}$ and
the set $\a=\{5,6,...,n-1\}$, then the generalized BCJ relation
(\ref{gen-BCJ}) can be written as following
\bea 0 & = & s_{\WH 1 234} A(\WH 1,2,3,4,\WH 5,6,...,n-1,n)+
\sum_{k=5}^{n-1}(s_{\WH 123}+ \sum_{i=1}^{k} s_{4i}) A(\WH 1,2,3,\WH
5,...,k,4,k+1,...,n)\nn
& & +  \sum_{5\leq k_1\leq  k_2\leq n-1}(s_{\WH 12}+ \sum_{i=1,i\neq
4}^{k_1} s_{3 i}+ \sum_{i=1}^{k_2}s_{4i}) A(\WH 1,2,\WH
5,...,k_1,3,...,k_2,4,...,n)\nn
& & + \sum_{5\leq k_1\leq  k_2\leq k_3\leq n-1}( \sum_{i=1,i\neq
3,4}^{k_1} s_{2 i}+\sum_{i=1,i\neq 4}^{k_2} s_{3 i}+
\sum_{i=1}^{k_3}s_{4i}) A(\WH 1,\WH
5,...,k_1,2,...,k_2,3,...,k_3,4,...,n)~~~\label{Non-BCJ}\eea
where it is important to notice that the sum like $\sum_{i=1}^{k}
s_{4i}$ is independent of $z$. Using (\ref{Non-BCJ}) we can solve
\bea & &  A(\WH 1,2,3,4,\WH 5,6,...,n-1,n)  =   T_1+ T_2+ T_3\nn
& & T_1  =  {s_{\WH 123}\over -s_{\WH 1234}}\sum_{k=5}^{n-1} A(\WH
1,2,3,\WH 5,...,k,4,k+1,...,n)+{s_{\WH 12}\over -s_{\WH
1234}}\sum_{5\leq k_1\leq  k_2\leq n-1} A(\WH 1,2,\WH
5,...,k_1,3,...,k_2,4,...,n)\nn
& & T_2 = {1\over -s_{\WH 1234}}\left\{\sum_{5\leq k_1\leq  k_2\leq
k_3\leq n-1}( \sum_{i=1,i\neq 3,4}^{k_1} s_{2 i}+\sum_{i=1,i\neq
4}^{k_2} s_{3 i}+ \sum_{i=1}^{k_3}s_{4i}) A(\WH 1,\WH
5,...,k_1,2,...,k_2,3,...,k_3,4,...,n)\right\}\nn
& & T_3  =  {1\over -s_{\WH 1234}}\left\{\sum_{k=5}^{n-1}(
\sum_{i=1}^{k} s_{4i}) A(\WH 1,2,3,\WH 5,...,k,4,k+1,...,n)\right.
\nn & &\left. +  \sum_{5\leq k_1\leq  k_2\leq n-1}( \sum_{i=1,i\neq
4}^{k_1} s_{3 i}+ \sum_{i=1}^{k_2}s_{4i}) A(\WH 1,2,\WH
5,...,k_1,3,...,k_2,4,...,n)\right\}~,~~~\label{Non-A-exp}\eea
where we have split all contributions into three types according to
their large-$z$ behavior. For the $T_3$ part, since $1,5$ are not
nearby and there is factor ${1\over -s_{\WH 1234}}$, the large-$z$
behavior will be $\xi_{1\mu}(z)\xi_{5\nu}(z){ G^{\mu\nu}(z)\over
z^{2}}$, which is  the prediction given in (\ref{Not-adj}), thus it
is safe to neglect this part. The naive large-$z$ behavior for $T_1,
T_2$ parts is $\xi_{1\mu}(z)\xi_{5\nu}(z){ G^{\mu\nu}(z)\over
z^{2}}$ and we need to investigate further.

At  the next step we  show that after iterations $T_1$ can be
reduced to the sum of forms ${a\over -s_{\WH 1234}} A(\WH 1, ...,\WH
5,...)$ and forms ${b\over -s_{\WH 1234}} A(\WH 1, \WH 5, ...)$
where $a,b$ are both independent of $z$. To see it,
 we will use similar generalized BCJ relation like
the one given in (\ref{Non-BCJ})  for $A(\WH 1,2,3,\WH
5,...,k,4,k+1,...,n)$ with the set $\b=\{2,3\}$ and $A(\WH 1,2,\WH
5,...,k_1,3,...,k_2,4,...,n)$ with the set $\b=\{2\}$, thus we can
solve
\bea A(\WH 1,2,\WH 5,...,k_1,3,...,k_2,4,...,n)= {1\over -s_{\WH
12}} \sum_t b_t A(\WH 1,\WH 5, ...,2,...,n)~\label{j=2-A12}\eea
 and similarly
\bea A(\WH 1,2,3,\WH 5,...,k,4,k+1,...,n) & = & {s_{\WH 1 2}\over
-s_{\WH 123}} A(\WH 1, 2, \WH 5,...,3,....)+ \sum { b\over  -s_{\WH
123}} A(\WH 1, \WH 5,....,2,...,3,...)\nn & & +\sum
 {a\over -s_{\WH 123}} A(\WH 1,2,\WH 5,...,3,....,n)\nn
 & = & \sum { b\over  -s_{\WH
123}} A(\WH 1, \WH 5,....,2,...,3,...) +\sum
 {a\over -s_{\WH 123}} A(\WH 1,2,\WH 5,...,3,....,n)~\label{j=2-A123}\eea
where the first term at the right hand side of first line in
(\ref{j=2-A123}) has been  reduced further using (\ref{j=2-A12}).
Putting (\ref{j=2-A123}) and (\ref{j=2-A12}) back to $T_1$, we see that $T_1$
reduces to the form we claimed.

Having understood $T_1$, Equation (\ref{Non-A-exp}) can be written
as the following sum
\bea A(\WH 1,2,3,4,\WH 5,6,...,n-1,n)  =\sum_t {a_t\over -s_{\WH
1234}} A(\WH 1, ...,\WH 5,...)+\sum_t {b_t\over -s_{\WH 1234}} A(\WH
1, \WH 5, ...)~~~\label{Non-A-exp-1}\eea
where the first part has the right large-$z$ behavior and can be
neglected. To show the conjecture (\ref{Not-adj}), our remaining task is to
show that the cyclic sum of the second part of (\ref{Non-A-exp-1})
is zero.

Now we work out the ${b_t\over -s_{\WH 1234}} A(\WH 1, \WH 5, ...)$
part for $A(\WH 1, \sigma_2,\sigma_3,\sigma_4,\WH 5, 6,...,n)$ where
$\sigma_{2,3,4}$ is reordering of $(2,3,4)$. First it is easy to see
that all $A$ will be the form that  $\sigma_{2,3,4}$ are inserted
between $5$ and $n$ while keeping the ordering of $(6,...,n-1)$. The
relative ordering of $\sigma_2,\sigma_3,\sigma_4$ can be arbitrary
and the $z$-independent factor $b$ is given as following:
\begin{itemize}

\item {\bf Type $A(\WH 1,\WH
5,...,k_1,\sigma_2,...,k_2,\sigma_3,...,k_3,\sigma_4,...,n)$:} This
type can come from several places. The first place is from the $T_2$
part of (\ref{Non-A-exp}) and the coefficient $b$ is given by
\bea b_I\left[\begin{array}{ccc} \sigma_2 & \sigma_3 & \sigma_4
\\ k_1 & k_2 & k_3 \end{array}\right]
=\sum_{i=1,i\neq \sigma_3,\sigma_4}^{k_1} s_{\sigma_2
i}+\sum_{i=1,i\neq \sigma_4}^{k_2} s_{\sigma_3 i}+
\sum_{i=1}^{k_3}s_{\sigma_4 i}\eea
where the first parameter $(\sigma_2,\sigma_3,\sigma_4)$ gives the
relative ordering of these three elements and the second parameter
$(k_1,k_2,k_3)$ tells which elements at the nearest left hand side
of corresponding $\sigma_i$. The second place is from the $A(\WH
1,\sigma_2,\WH 5,...,n)$ in $T_1$ of (\ref{Non-A-exp}) with another
solving using the generalized BCJ relation as given in
(\ref{j=2-A12}). To make it clear, we say that the path of second
contribution is
\bea {\rm Path}_{II}\equiv (\WH 1,\sigma_2,\sigma_3,\sigma_4,\WH
5)\to (\WH 1,\sigma_2,\WH 5, \sigma_3,\sigma_4)\to (\WH 1, \WH 5,
\sigma_3,\sigma_4)\eea
while the path of the first contribution is
\bea {\rm Path}_{I}\equiv (\WH 1,\sigma_2,\sigma_3,\sigma_4,\WH
5)\to  (\WH 1, \WH 5, \sigma_2,\sigma_3,\sigma_4)\eea
The corresponding coefficient of second path is
\bea  b_{II}\left[\begin{array}{ccc} \sigma_2 & \sigma_3 & \sigma_4
\\ k_1 & k_2 & k_3 \end{array}\right]= -\sum_{i=1,i\neq \sigma_3,\sigma_4}^{k_1} s_{\sigma_2
i}\eea
The third contribution comes from the path
\bea {\rm Path}_{III}\equiv (\WH 1,\sigma_2,\sigma_3,\sigma_4,\WH
5)\to (\WH 1,\sigma_2,\sigma_3,\WH 5)\to (\WH 1, \WH 5,
\sigma_2,\sigma_3)\eea
and is given by
\bea  b_{III}\left[\begin{array}{ccc} \sigma_2 & \sigma_3 & \sigma_4
\\ k_1 & k_2 & k_3 \end{array}\right]= -\sum_{i=1,i\neq \sigma_3,\sigma_4}^{k_1} s_{\sigma_2
i}-\sum_{i=1,i\neq \sigma_4}^{k_2} s_{\sigma_3 i}\eea
The fourth contribution comes from the path
\bea {\rm Path}_{IV}\equiv (\WH 1,\sigma_2,\sigma_3,\sigma_4,\WH
5)\to (\WH 1,\sigma_2,\sigma_3,\WH 5)\to (\WH 1, \sigma_2,\WH 5)\to
(\WH 1, \WH 5)\eea
and is given by
\bea  b_{IV}\left[\begin{array}{ccc} \sigma_2 & \sigma_3 & \sigma_4
\\ k_1 & k_2 & k_3 \end{array}\right]= +\sum_{i=1,i\neq \sigma_3,\sigma_4}^{k_1} s_{\sigma_2
i}\eea
Summing these four contribution together we have
\bea b^{(\sigma_2,\sigma_3,\sigma_4)}\left[\begin{array}{ccc}
\sigma_2 & \sigma_3 & \sigma_4
\\ k_1 & k_2 & k_3 \end{array}\right]=
\sum_{i=1}^{k_3}s_{\sigma_4 i} ~~~\label{Type-I}\eea
where the superscript tells that the original amplitude is the
ordering $A(\WH 1,\sigma_2,\sigma_3,\sigma_4,\WH 5,...,n)$.

\item {\bf Type $A(\WH 1,\WH
5,...,k_1,\sigma_3,...,k_2,\sigma_2,...,k_3,\sigma_4,...,n)$:} There
are several pathes giving contributions. The first one is from path
${\rm Path}_{II}$ as
\bea  b_{II}\left[\begin{array}{ccc} \sigma_3 & \sigma_2 & \sigma_4
\\ k_1 & k_2 & k_3 \end{array}\right]= -\sum_{i=1,i\neq \sigma_3}^{k_2} s_{\sigma_2 i}\eea
The second contribution is from the fourth path as
\bea  b_{IV}\left[\begin{array}{ccc} \sigma_3 & \sigma_2 & \sigma_4
\\ k_1 & k_2 & k_3 \end{array}\right]= \sum_{i=1,i\neq \sigma_3}^{k_2} s_{\sigma_2 i}\eea
Adding them up we get zero for this type.

\item {\bf Type $A(\WH 1,\WH
5,...,k_1,\sigma_3,...,k_2,\sigma_4,...,k_3,\sigma_2,...,n)$:}
Contributions from various pathes are given as
\bea  b_{II}\left[\begin{array}{ccc} \sigma_3 & \sigma_4 & \sigma_2
\\ k_1 & k_2 & k_3 \end{array}\right]= -\sum_{i=1}^{k_3} s_{\sigma_2 i}\nn
 b_{IV}\left[\begin{array}{ccc} \sigma_3 & \sigma_4 & \sigma_2
\\ k_1 & k_2 & k_3 \end{array}\right]= +\sum_{i=1}^{k_3} s_{\sigma_2 i}\eea
thus the total contribution is zero.

\item {\bf Type $A(\WH 1,\WH
5,...,k_1,\sigma_2,...,k_2,\sigma_4,...,k_3,\sigma_3,...,n)$:}
Contributions from various pathes are given as
\bea b_{III}\left[\begin{array}{ccc} \sigma_2 & \sigma_4 & \sigma_3
\\ k_1 & k_2 & k_3 \end{array}\right] &= & -\sum_{i=1,i\neq \sigma_3,\sigma_4}^{k_1} s_{\sigma_2
i}-\sum_{i=1}^{k_3} s_{\sigma_3 i}\nn
b_{IV}\left[\begin{array}{ccc} \sigma_2 & \sigma_4 & \sigma_3
\\ k_1 & k_2 & k_3 \end{array}\right] &= & +\sum_{i=1,i\neq \sigma_3,\sigma_4}^{k_1} s_{\sigma_2
i}\eea
Putting together we have
\bea b^{(\sigma_2,\sigma_3,\sigma_4)}\left[\begin{array}{ccc}
\sigma_2 & \sigma_4 & \sigma_3
\\ k_1 & k_2 & k_3 \end{array}\right] &= &-\sum_{i=1}^{k_3} s_{\sigma_3 i}~~~\label{Type-II} \eea

\item {\bf Type $A(\WH 1,\WH
5,...,k_1,\sigma_4,...,k_2,\sigma_2,...,k_3,\sigma_3,...,n)$:}
Contributions from various pathes are given as
\bea  b_{III}\left[\begin{array}{ccc} \sigma_4 & \sigma_2 & \sigma_3
\\ k_1 & k_2 & k_3 \end{array}\right] &= & -\sum_{i=1,i\neq \sigma_3}^{k_2} s_{\sigma_2
i}-\sum_{i=1}^{k_3} s_{\sigma_3 i}\nn
b_{IV}\left[\begin{array}{ccc} \sigma_4 & \sigma_2 & \sigma_3
\\ k_1 & k_2 & k_3 \end{array}\right] &= & +\sum_{i=1,i\neq \sigma_3}^{k_2} s_{\sigma_2
i}\eea
Putting together we have
\bea b^{(\sigma_2,\sigma_3,\sigma_4)}\left[\begin{array}{ccc}
\sigma_4 & \sigma_2 & \sigma_3
\\ k_1 & k_2 & k_3 \end{array}\right] &= & -\sum_{i=1}^{k_3} s_{\sigma_3 i}~~~\label{Type-III}\eea

\item {\bf Type $A(\WH 1,\WH
5,...,k_1,\sigma_4,...,k_2,\sigma_3,...,k_3,\sigma_2,...,n)$:} Only
the fourth path gives nonzero contribution, thus we have
\bea  b^{(\sigma_2,\sigma_3,\sigma_4)}\left[\begin{array}{ccc}
\sigma_4 & \sigma_3 & \sigma_2
\\ k_1 & k_2 & k_3 \end{array}\right] &= & +\sum_{i=1}^{k_3} s_{\sigma_2
i}~~~\label{Type-IV} \eea

\end{itemize}

Having established (\ref{Type-I}), (\ref{Type-II}), (\ref{Type-III})
and (\ref{Type-IV}), we now show the cyclic sum of all $b$
coefficients is zero by listing out following table (\ref{j=5-exa}):

\bea
\begin{array}{ccccc}  & ~~A(\WH 1,2,3,4, \WH 5)~~ & ~~A(\WH 1,3,4,2, \WH
5)~~
& ~~A(\WH 1,4, 2,3, \WH 5)~~ & {\rm cyclic~sum} \\
A(\WH 1,\WH 5, k_1,2,k_2,3,k_3,4) & +\sum_{i=1}^{k_3} s_{4 i}& -\sum_{i=1}^{k_3} s_{4i} &  0 & 0 \\
A(\WH 1,\WH 5, k_1,2,k_2,4,k_3,3) & -\sum_{i=1}^{k_3} s_{3i}& +\sum_{i=1}^{k_3} s_{3i} & 0& 0\\
A(\WH 1,\WH 5, k_1,3,k_2,2,k_3,4) & 0 & -\sum_{i=1}^{k_3} s_{4i}&+\sum_{i=1}^{k_3} s_{4i}& 0\\
A(\WH 1,\WH 5, k_1,3,k_2,4,k_3,2) & 0 &  +\sum_{i=1}^{k_3} s_{2i} & -\sum_{i=1}^{k_3} s_{2i}& 0\\
A(\WH 1,\WH 5, k_1,4,k_2,2,k_3,3) & -\sum_{i=1}^{k_3} s_{3i}& 0 &  +\sum_{i=1}^{k_3} s_{3i}& 0\\
A(\WH 1,\WH 5, k_1,4,k_2,3,k_3,2) & +\sum_{i=1}^{k_3} s_{2i} & 0 &
-\sum_{i=1}^{k_3} s_{2i}& 0
\end{array}~~~\label{j=5-exa}\eea
%

 \subsection{A general proof}

Having above example for $j=5$, now we give the proof for general
$j$. The idea of the proof is following. First we use the generalized
BCJ relation to write $A(\WH 1, \sigma(2,...,j-1),\WH j, j+1,...,n)$
as the sum of the form $-\frac{1}{s_{\WH1...,i-1}}a A(\WH1,...,\WH
i,...)$ and the form $-\frac{1}{s_{\WH1...,i-1}}b A(\WH1,\WH i,...)$
with $z$-independent coefficients $a,b$. Terms with form
$-\frac{1}{s_{\WH1...,i-1}}a A(\WH1,...,\WH i,...)$ have the right
large-$z$ behavior and they are safe to be neglected. Terms with
form $-\frac{1}{s_{\WH1...,i-1}}b A(\WH1,\WH i,...)$ are dangerous,
so we need to show that after the cyclic sum these contributions are
zero.

To do so we need to find the expression for coefficient $b$. To see
the pattern we rewrite results (\ref{Type-I}), (\ref{Type-II}),
(\ref{Type-III}) and (\ref{Type-IV}) as
\bea &&A_{lead}(\WH1,2,3,4,\WH5,...,n)\nn
&=&-\frac{1}{s_{\WH1234}}\Sl_{\sigma\in \W P(O\{2,3,4\}\bigcup
\emptyset^T)}\left(\Sl_{\rho\in P(O\{\sigma\}\bigcup\{6,...,n\})
}S_4(\rho)A(\WH1,\WH5,\rho,n)\right)\nn
&&+\frac{1}{s_{\WH1234}}\Sl_{\sigma\in \W P(O\{2,3\}\bigcup
O^T\{4\})}\left(\Sl_{\rho\in
P(O\{\sigma\}\bigcup\{6,...,n\})}S_3(\rho)A(\WH1,\WH5,\rho,n)\right)\nn
&&-\frac{1}{s_{\WH1234}}\Sl_{\sigma\in \W P(O\{2\}\bigcup
O^T\{3,4\})}\left(\Sl_{\rho\in
P(O\{\sigma\}\bigcup\{6,...,n\})}S_2(\rho)A(\WH1,\WH5,\rho,n)\right).~~~\label{S-appear}
\eea
where $ O^T(\a)$ means the reversed ordering of set $\a$ and
$S_j(\rho)$ means the sum of $s_{ji}$ for each element $i$ in
the ordering of all external legs(in this case the ordering $\WH1$, $\WH5$, $\rho$, $n$)
at the left hand side of element $j$ in the given ordering $\rho$. The
constraint permutations $\W P(\a\bigcup \b)$ means all permutations
satisfying following three conditions: (1) relative ordering of
elements in the set $\a$ is kept; (2) relative ordering of elements
in the set $\b$ is kept; (3) the last elements is always the last
element of the set $\a$.

Having above observation, we can write down the general pattern as
\bea &&A_{lead}(\WH1,2,3,4,...,i-1, \WH
i,...,n)~~~~\label{gen-lead}\\
& = & -\frac{1}{s_{\WH12,...,i-1}} \sum_{j=2}^{i-1}\Sl_{\sigma\in \W
P(O\{2,...,j\}\bigcup O^T\{j+1,...,i-1\})} (-)^{i-1-j}\left(
\Sl_{\rho\in P(O\{\sigma\}\bigcup O\{i+1,...,n-1\})} S_j(\rho)
A(\WH1,\WH i,\rho,n)\right)\nonumber \eea
Assuming this pattern is right, now we show the cyclic sum is zero,
i.e., coefficient of  a given ordering of $\rho$ is zero. To do so,
first we need to find where the ordering $\rho$ can appear in the
pattern (\ref{gen-lead}).  Since $\rho\in P(O\{\sigma\}\bigcup
O\{i+1,...,n-1\})$, we see that there is one and only one ordering
of $\sigma$ can give the $\rho$. For given ordering $\sigma$, since
$\sigma\in \W P(O\{\a\}\bigcup O^T\{\b\})$,  the last element of set
$\a$ is determined by the last element of ordering $\sigma$.
Similarly, the first element of ordering $\sigma$ must be the first
element $\a_{first}=\sigma_{first}$ of set $\a$ or the last element
$\b_{last}=\sigma_{first}$ of the set $\b$. Because the amplitude is
given by $A(\WH 1, \a \bigcup \b,\WH i, i+1,...,n)$, we see that
there are two and only two amplitudes in the cyclic sum can give
contributions to ordering $\rho$ and they are $A(\WH 1, \{
\sigma_{first},\W \a,\sigma_{last}\} \bigcup \{ \b\},\WH i,...,n)$
and $A(\WH 1, \{ \W \a,\sigma_{last}\} \bigcup \{ \b,
\sigma_{first}\},\WH i,...,n)$. Now from the general pattern
(\ref{gen-lead}) it is easy to see that these two contributions are
same with opposite signs and their sum is zero.

Now let us prove the pattern \eqref{gen-lead}. Using the generalized
BCJ relation \eqref{gen-BCJ}  (like the one given in
(\ref{Non-A-exp})) and the leading part \eqref{gen-lead} for
amplitudes with fewer legs between $1$ and $i$, we can express the
leading part of the amplitude $A(\WH1,2,3,...,i-1,\WH i,...,n)$ as
\bea &&A_{lead}(\WH1,2,3,...,i-1,\WH i,...,n)\nn
&=&-\frac{1}{s_{\WH12...i-1}}\Biggl[\Sl_{\rho\in
P(O\{2,...,i-1\}\bigcup
O\{i+1,...,n-1\})}(S_{2}(\rho)+...+S_{i-1}(\rho))A(\WH1,\WH
i,\rho,n)~~~\label{A-lead-rec} \nn
&+& \Sl_{k=2}^{i-2} \Sl_{j=2}^{k}(-1)^{k+1-j} \Sl_{\sigma\in \W
P(O\{2,...,j\}\bigcup O^T\{j+1,...,k\})} \Sl_{\rho\in P(O\{\sigma\}\bigcup
O\{k+1,...,i-1\}\bigcup  O\{i+1,...,n-1\}))} S_j(\rho)A(\WH1,\WH
i,\rho,n) \Biggr].\nn \eea
We define $\xi_{\rho(l)}$ to be  the position of the  $l$-th
external particles  in the permutation $\rho$. It is worth to notice
that in the last line of (\ref{A-lead-rec}), although the relative
ordering inside each set $\sigma$, $\{k+1,...,i-1\}$ and
$\{i+1,...,n-1\}$ is kept, these three sets can have arbitrary
relative ordering and we will divide relative ordering among three
sets in following discussions.


Let us start from  the ordering $\xi_{\rho(2)}<
\xi_{\rho(3)}<...<\xi_{\rho(i-1)}$, which can come from the second
line and the third line with $j=k$ in equation \eqref{A-lead-rec}.
The contribution from the second line is given by
\bea -\Sl_{\rho\in P(O\{2,...,i-1\}\bigcup
O\{i+1,...,n-1\})}\frac{1}{s_{\WH12...i-1}}(S_{2}(\rho)+...+S_{i-1}(\rho))A(\WH1,\WH
i,\rho,n), \eea
while contributions from the third line with $j=k$ are given by
\bea \Sl_{\rho\in P(O\{2,...,i-1\}\bigcup
O\{i+1,...,n-1\})}\frac{1}{s_{\WH12...i-1}}S_{k}(\rho))A(\WH1,\WH
i,\rho,n),~~~2\leq k\leq i-2\eea
Summing over $k$ and the second line contributions together we have
\bea -\frac{1}{s_{\WH12...i-1}}\Sl_{\rho\in P(O\{2,...,i-1\}\bigcup
O\{i+1,...,n-1\})}S_{i-1}(\rho)A(\WH1,\WH i,\rho,n).~~~\label{l=2}
\eea
This is
noting but just the $j=i-1$ part of \eqref{gen-lead}.

Now we consider all permutations with following condition:
$\xi_{\rho(i-1)}> \xi_{\rho(i-2)}>...> \xi_{\rho(l)}$, but
$\xi_{\rho(l-1)}>\xi_{\rho(l)}$ for a given $l$. In other words, $l$ is the first
one breaks the natural descendent ordering from $i-1$ to $2$. It is
easy to see that the second line and the third line with $k<l-1$ in
\eqref{A-lead-rec} can not give such permutations. When $k>l$, from
$\sigma\in \W P(O\{2,...,j\}\bigcup O^T\{j+1,...,k\})$ (especially the
$O^T\{j+1,...,k\}$ part) we see that there is no contribution
either. There are only two contributions coming from $k=l-1$ and
$k=l$.

When $k=l-1$, all $2\leq j\leq l-1$ will contribute to this
ordering. For a given $j$, we get
\bea
-\frac{1}{s_{\WH12...i-1}}(-1)^{l-j}\Sl_{\sigma\in \W
P(O\{2,...,j\}\bigcup O^T\{j+1,...,l-1\})}\Sl_{\rho\in P(O\{\sigma\}\bigcup
O\{l,...,i-1\}\bigcup O\{i+1,...,n-1\}))}S_{j}(\rho)A(\WH1,\WH i,\rho,n).\nonumber
\eea
When $k=l$, only $2\leq j\leq l-1$ contribute to this ordering and
$j=l$ does no contribute. For a given $j$, we get
 \bea &&-\frac{1}{s_{\WH12...i-1}}(-1)^{l+1-j}\Sl_{\sigma\in \W
P(O\{2,...,j\}\bigcup O^T\{j+1,...,l-1,l\})}\Sl_{\rho\in P(O\{\sigma\}\bigcup
O\{l+1,...,i-1\}\bigcup O\{i+1,...,n-1\})}S_{j}(\rho)A(\WH1,\WH i,\rho,n)\nn
&\to&-\frac{1}{s_{\WH12...i-1}}(-1)^{l+1-j}\Sl_{\sigma\in \W
P(O\{2,...,j\}\bigcup O^T\{j+1,...,l-1\})}\Sl_{\rho\in
P(O\{\sigma\}\bigcup O\{l,l+1,...,i-1\}\bigcup
O\{i+1,...,n-1\})}S_{j}(\rho)A(\WH1,\WH i,\rho,n).\nonumber \eea
where because we have required that $\xi_{\rho(l)}<\xi_{\rho(l-1)}$
and $\xi_{\rho(l)}<\xi_{\rho(l+1)}$, the first line is equivalent to
the second line. Thus for any given $j$ ($2\leq j\leq l-1$),
contributions from  cases $k=l-1$ and $k=l$ will cancel each other.

Above cancelation will work for $2<l<i-1$. The case $l=2$ has been
discussed in \eqref{l=2}. For the case $l=i-1$, i.e.,
$\xi_{\rho(i-2)}>\xi_{\rho(i-1)}$, it can only come from the third
line of \eqref{A-lead-rec} with $k=i-2$, thus we have
\bea
&&-\frac{1}{s_{\WH12...i-1}}\Sl_{j=2}^{i-2}(-1)^{i-1-j}\Sl_{\sigma\in
\W P(O\{2,...,j\}\bigcup O^T\{j+1,...,i-2\})}\Sl_{\rho\in
P(\{\sigma\}\bigcup O\{O\{i-1\}\bigcup
O\{i+1,...,n-1\})\})}S_j(\rho)A(\WH1,\WH i,\rho,n)\nn
&\to
&-\frac{1}{s_{\WH12...i-1}}\Sl_{j=2}^{i-2}(-1)^{i-1-j}\Sl_{\sigma\in\W
P(O\{2,...,j\}\bigcup O^T\{j+1,...,i-2,i-1\})}\Sl_{\rho\in
P(\{\sigma\}\bigcup O\{i+1,...,n-1\})}S_j(\rho)A(\WH1,\WH
i,\rho,n).~~~\label{l=i-1} \eea
where again the first line is equivalent to the second line because
we have required that $\xi_{\rho(i-2)}>\xi_{\rho(i-1)}$.

After summing the contributions from \eqref{l=2} and \eqref{l=i-1},
we get the general pattern \eqref{gen-lead}. From above proof, we
can see that the study of cyclic sum is much more difficult than the
study of permutation sum discussed in \cite{Du:2011se}.

\section{The adjacent case of cyclic sum}

Having proved the conjecture (\ref{Boels-2}) for non-adjacent case,
we move to the adjacent case, which can happen when and only when
the shifted pair is $(1,n)$. In this case, the boundary behavior is
\bea \sum_{cyclic~\a} A_n (\WH 1, \{\a(2,...,n-1)\}, \WH n)\to
\xi_{1\mu}(z)\xi_{n\nu}(z){ G^{\mu\nu}(z)\over
z}~~~~\label{B-adj}\eea
Comparing to the proof of cyclic sum of non-adjacent case, adjacent
case is much more simpler and essentially only the KK-relations are
needed. Again we will use one example to demonstrate our idea of
proof and then give the general proof.

\subsection{Five point example}

We consider the five point case with  $(1,5)$-deformation. In this
case, we have three amplitudes $A(\WH 1,2,3,4,\WH5)$, $A(\WH
1,3,4,2,\WH5)$ and $A(\WH 1,4,2,3,\WH5)$. For these amplitudes, the
first step is to use KK relation to write an amplitudes in terms of
amplitudes with $4$, $5$ adjacent, thus we have
\bea A(\WH 5,\{\WH 1, 3\}, 4, \{2\})& = & -A(\WH 5,\WH 1, 3,
2,4)-A(\WH 5,\WH 1,2, 3, 4)-A(\WH 5,2,\WH 1, 3, 4)  \nn & = &
-A(\WH1,2,3,4,\WH5)-A(\WH1,3,2,4,\WH5)-A(2,\WH1,3,4,\WH5),~~~\label{adj-5-1}\eea
and
\bea A(\WH 5, \{\WH 1\}, 4 ,\{2,3\})& = &  A(\WH 5, \WH 1, 3,2,
4)+A(\WH 5, 3,\WH 1,2,  4)+A(\WH 5, 3,2, \WH 1 4)\nn & =&
A(\WH1,3,2,4,\WH5)+A(3,\WH1,2,4,\WH5)+A(3,2,\WH1,4,\WH5).
~~~\label{adj-5-2}\eea
Having the expansion (\ref{adj-5-1}) and  (\ref{adj-5-2}), we can
observe that all six terms at the right hand side can be divided
into following two types. The first type is the form
 $A(\WH1,...,4,\WH5)$ where $\WH 1, \WH 5$ are adjacent and
second type is the form $A(...,\WH1,...,4,\WH 5)$ where $\WH 1,\WH
5$ are not adjacent. For the second type, since $\WH 1, \WH 5$ are
not adjacent, each amplitude has the boundary behavior
$\xi_{1\mu}(z)\xi_{5\nu}(z){ G^{\mu\nu}(z)\over z}$, so it is the
boundary behavior we try to prove for (\ref{B-adj}).

For the first type, although the boundary behavior of each amplitude
is worse than the wanted (\ref{B-adj}) since $\WH 1, \WH 5$ are
adjacent, their sum is, in fact, zero which can be easily seen from
\bea A(\WH
1,2,3,4,\WH5)+[-A(\WH1,2,3,4,\WH5)-A(\WH1,3,2,4,\WH5)]+[A(\WH1,3,2,4,\WH5)]=0\eea
%

\subsection{A general proof}

Having above example, now we give a general proof for the adjacent
case. The cyclic sum will be given by
\bea I=A(\WH 1, 2,...,n-1, \WH n)+\sum_{i=2}^{n-2}  A(\WH 1,
i+1,...,n-1,2,...,i,\WH n) ~~~\label{adj-I-sum}\eea
where in the first term, $n-1, n$ are nearby while in other terms,
they are not. As in demonstrated example, the first step is to use
KK-relation to expand other terms in the form with $n-1, n$ nearby
\bea  A(\WH 1, i+1,...,n-1,2,...,i,\WH n) & = & (-)^{i-1}
\sum_{\sigma \in P(O(1,i+1,i+2,...,n-2)\bigcup O(i,i-1,...,2))}
A(\WH n, \sigma , n-1)~.~~\label{adj-exp-1} \eea
Among all terms in (\ref{adj-exp-1}), some will have the form $A(\WH
n,...,\WH 1,...,n-1)$ which will give the wanted large-$z$ behavior,
while other terms, which we will call the leading part, will have
the form $A(\WH n, \WH 1, ..., n-1)$. These leading terms can be
written as
\bea  A_{lead}(\WH 1, i+1,...,n-1,2,...,i,\WH n) & = & (-)^{i-1}
\sum_{\sigma \in P(O(i+1,i+2,...,n-2)\bigcup O(i,i-1,...,2))} A( \WH
1, \sigma , n-1,\WH n)~.~~\label{adj-exp-lead} \eea
where when $i=n-2$, (\ref{adj-exp-lead}) is reduced to just
$(-)^{n-3} A(\WH 1,n-2,n-3,...,2,n-1,\WH n)$. Putting
(\ref{adj-exp-lead}) back to (\ref{adj-I-sum}) we have
\bea I_{lead}& = & A(\WH 1, 2,...,n-1, \WH
n)+\sum_{i=2}^{n-3}(-)^{i-1} \sum_{\sigma \in
P(O(i+1,i+2,...,n-2)\bigcup O(i,i-1,...,2))} A( \WH 1, \sigma ,
n-1,\WH n) \nn
& & +(-)^{n-3} A(\WH 1,n-2,n-3,...,2,n-1,\WH n)
~~~\label{adj-I-sum}\eea
To show $I_{lead}=0$, it is important to notice that
$P(O(i+1,i+2,...,n-2)\bigcup O(i,i-1,...,2))$ means that either
$i+1$ or $i$ at the second position, i.e.,
\bean & & \sum_{\sigma \in P(O(i+1,i+2,...,n-2)\bigcup
O(i,i-1,...,2))} A( \WH 1, \sigma , n-1,\WH n)\nn
& = & \sum_{\sigma \in P(O(i+1,i+2,...,n-2)\bigcup O(i-1,...,2))} A(
\WH 1,i, \sigma , n-1,\WH n)\nn & & + \sum_{\sigma \in
P(O(i+2,...,n-2)\bigcup O(i,i-1,...,2))} A( \WH 1, i+1, \sigma ,
n-1,\WH n) ~. \eean
Using this observation we have
\bean I_{lead}& = & A(\WH 1, 2,...,n-1, \WH n)\nn & & + (-) A(\WH
1,2,3,...,n-1,\WH n)+ \sum_{i=3}^{n-3}(-)^{i-1} \sum_{\sigma \in
P(O(i+1,i+2,...,n-2)\bigcup O(i-1,...,2))} A( \WH 1,i, \sigma ,
n-1,\WH n)\nn & & + \sum_{i=2}^{n-4}(-)^{i-1} \sum_{\sigma \in
P(O(i+2,...,n-2)\bigcup O(i,i-1,...,2))} A( \WH 1, i+1, \sigma ,
n-1,\WH n)+(-)^{n-4} A(\WH 1,n-2,...,2,n-1,\WH n)\nn & & +(-)^{n-3}
A(\WH 1,n-2,n-3,...,2,n-1,\WH n)\eean
where  we have split the $i=2$ term from the summation at the second
line and $i=n-3$ term from the summation at the third line. With
above rewriting, it is easy to see the cancelation of the summation
at the second line and third line, so $I_{lead}=0$.

Having shown $I_{lead}=0$, we know that
\bea I & = & \sum_{\a,\b,~not~empty} c_i A(\WH 1, \a, \WH n, \b)~.
\eea
Because for each term the $1,n$ are not nearby, thus the large-$z$
behavior is
\bea I\sim  A(\WH 1, \a, \WH n, \b) \sim \xi_{1\mu}(z)\xi_{n\nu}(z){
G^{\mu\nu}(z)\over z}\eea
and we have finished the proof.

\section{The partial-ordered  permutation sum}
In this section, we will investigate the large $z$-behavior of
following expression
\bea I_{partial}\equiv \Sl_{\sigma\in
P(O\{2,...,l\}\bigcup\{l+1,...,i-1\})}A(\WH1,\sigma,\WH i,...,n)
~~~\label{I-par-def}\eea
where the sum is over all permutations of elements $\{2,3,...,i-1\}$
under one condition: the relative ordering in the subset
$\{2,...,l\}$ is kept. We will show that its large $z$-behavior is
like
\bea I_{partial} \sim \frac{1}{z}\Sl_{\sigma'\in
P(\{l+1,...,i-1\})}A(\WH1,\sigma',\WH
i,...,n)~~~\label{ordered-perm} \eea
where $1,i$ are not adjacent. For adjacent case, i.e., the
partial-ordered permutation sum is given by
 \bea I_{partial}^{adj}\equiv
\Sl_{\sigma\in
P(O\{2,...,l\}\bigcup\{l+1,...,n-1\})}A(\WH1,\sigma,\WH n), \eea
where both sets $\{2,...,l\}$ and $\{2,...,l\}$ are not empty, using
the $U(1)$-decoupling identity, one can write the above expression
as
\bea I_{partial}^{adj}&=&-\Sl_{\sigma\in
P(O\{2,...,l\}\bigcup\{l+1,...,n-2\})}A(\WH1,\sigma,\WH n,n-1) \eea
Thus adjacent case is reduced to the nonadjacent case presented in
\eqref{ordered-perm}. Relations \eqref{ordered-perm} are new results
of our paper and will be applied to the proof of \eref{Boels-com}.

Result \eqref{ordered-perm} contains following special cases:
\begin{itemize}

\item When the set $\{l+1,...,i-1\}$ is empty, we have the familiar
result that the large $z$-behavior with non-adjacent deformation is
one power of $z$ better than the one with adjacent deformation
 \bea A(\WH1,...,\WH i,...,n)\sim \frac{1}{z}A(\WH1,\WH
i,...,n). \eea

\item When the set $\{2,...,l\}$ has only one element, the $\sigma\in
P(O\{2\}\bigcup\{3,...,i-1\}$ is nothing, but $\sigma\in
P(\{2,3,...,i-1\}$. Thus we reproduced the large $z$-behavior of
permutation sum. Because this, we will assume that the first set has
at least two elements.

\end{itemize}
These two special cases can be taken as the staring point of our
inductive proof. Before giving the general proof, we will present
two examples, where idea of our proof will be easier to understand.

\subsection{First example}
We consider a simple example with $O\{2,3\}$ as the ordered set and
$\{4\}$ as the permuted set
\bea T\equiv
A(\WH1,2,3,4,\WH5,...,n)+A(\WH1,2,4,3,\WH5,...,n)+A(\WH1,4,2,3,\WH5,...,n).
\eea
To see the $z$-dependence of $T$, we  write down  following two
relations. The first one is the generalized BCJ relation
\eqref{gen-BCJ} with set $\b=\{2,3\}$
 \bea
&&s_{\WH123}A(\WH1,2,3,4,\WH5,...,n)+(s_{\WH123}+s_{34})A(\WH1,2,4,3,\WH5,...,n)
+(s_{\WH123}+s_{34}+s_{24})A(\WH1,4,2,3,\WH5,...,n)\nn
&=&-\Sl_{\sigma\in P(\{3\}\bigcup
O\{6,...,n-1\})}\left(s_{\WH12}+S_3(\sigma)\right)A(\WH1,2,4,\WH5,\sigma,n)\nn
&&-\Sl_{\sigma\in P(\{3\}\bigcup
O\{6,...,n-1\})}\left(s_{\WH12}+s_{\WH24}
+S_3(\sigma)\right)A(\WH1,4,2,\WH5,\sigma,n)\nn
&&-\Sl_{\sigma\in P(\{2,3\}\bigcup
O\{6,...,n-1\})}\left(S_{2}(\sigma)+S_3(\sigma)\right)A(\WH1,4,\WH5,\sigma,n),
\eea
while the second one is the fundamental BCJ relation for elements in
the second set (here is just element $4$)
 \bea
&&s_{4\WH1}A(\WH1,4,2,3,\WH5,...,n)+(s_{4\WH1}+s_{42})A(\WH1,2,4,3,\WH5,...,n)
+(s_{4\WH1}+s_{42}+s_{43})A(\WH1,2,3,4,\WH5,...,n)\nn
&=&-\Sl_{\sigma''\in
P(\{4\}\bigcup\{6,...,n-1\})}S_4(\sigma)A(\WH1,2,3,\WH5,\sigma,n). \eea
Here notation $S_{l}(\sigma)$ has been defined under equation
\eqref{S-appear}. Summing these two  relations together, we get
\bea
&&A(\WH1,2,3,4,\WH5,...,n)+A(\WH1,2,4,3,\WH5,...,n)+A(\WH1,4,2,3,\WH5,...,n)\nn
&=&-\frac{1}{s_{\WH1234}}\Sl_{\sigma\in P(\{3\}\bigcup O\{6,...,n-1\})}
\left(s_{\WH12}+S_3(\sigma)\right)A(\WH1,2,4,\WH5,\sigma,n)\nn
&&-\frac{1}{s_{\WH1234}}\Sl_{\sigma\in P(\{3\}\bigcup O\{6,...,n-1\})}
\left(s_{\WH12}+s_{24}+S_3(\sigma)\right)A(\WH1,4,2,\WH5,\sigma,n)\nn
&&-\frac{1}{s_{\WH1234}}\Sl_{\sigma\in P(\{2,3\}\bigcup
O\{6,...,n-1\})}
\left(S_{2}(\sigma)+S_3(\sigma)\right)A(\WH1,4,\WH5,\sigma,n)\nn
&&-\frac{1}{s_{\WH1234}}\Sl_{\sigma\in P(O\{4\}\bigcup
O\{6,...,n-1\})} S_4(\sigma)A(\WH1,2,3,\WH5,\sigma,n). \eea
The sum at the right-handed side can be divided into two types. The
first type is
\bea -\frac{s_{\WH 1 2}}{s_{\WH1234}}\Sl_{\sigma\in P(\{3\}\bigcup
O\{6,...,n-1\})}
[A(\WH1,2,4,\WH5,\sigma,n)+A(\WH1,4,2,\WH5,\sigma,n)]\sim {1\over z}
A(\WH 1, 4,\WH 5,...)\eea
where we have used the result for $\sigma\in P(O\{ 2\}\bigcup\{4\})$
of \eqref{I-par-def}. The second type is remaining terms
\bea -\frac{1}{s_{\WH1234}} A(\WH 1, ..., \WH 5,...)\sim {1\over z}
A(\WH 1, 4,\WH 5,...)~.\eea
Thus we have shown \eqref{ordered-perm} for this example.

\subsection{Second example}

Previous example is a little bit simple. To see more clear the
pattern of general proof,  we consider another example
\bea T\equiv
\Sl_{\sigma\in P(O\{2,3,4\}\bigcup\{5,6\})}A(\WH1,\sigma,\WH
7,...,n). \eea
Now we consider two types of relations. The first type is
generalized BCJ relation with set  $\{\beta\}=\{2,3,4\}$ for each
ordering of permutations $\{ 5, 6\}$. For example, with ordering
$5,6$ we have
 \bea
0&=&\Sl_{\sigma\in P(O\{2,3,4\}\bigcup O\{5,6\})}\left(\WH
S_2(\sigma)+\WH S_3(\sigma)+\WH
S_4(\sigma)\right)A(\WH1,\sigma,\WH7,...,n)\nn
&+&\Sl_{\gamma\in P(O\{4\}\bigcup O\{8,...,n-1\})}\Sl_{\sigma\in
P(O\{2,3\}\bigcup O\{5,6\})}\left(\WH S_2(\sigma)+\WH
S_3(\sigma)+S_4(\gamma)\right)A(\WH1,\sigma,\WH7,\gamma,n)\nn
&+&\Sl_{\gamma\in P(O\{3,4\}\bigcup O\{8,...,n-1\})}\left(\WH
S_2(\sigma)+S_3(\gamma)+S_4(\gamma)\right)\Sl_{\sigma\in P(O\{2\}\bigcup
O\{5,6\})}A(\WH1,\sigma,\WH7,\gamma,n)\nn
&+&\Sl_{\gamma\in P(O\{2,3,4\}\bigcup
O\{8,...,n-1\})}\left(S_2(\gamma)+S_3(\gamma)+S_4(\gamma)\right)\Sl_{\sigma\in
P( O\{5,6\})}A(\WH1,\sigma,\WH7,\gamma,n) ~~~\label{E2-56}\eea
where to distinguish the $z$-dependence, we have used $\WH
S_2(\sigma)$ to mean that  it is $s_{2\WH 1}$  in the sum while for $
S_2(\gamma)$,  it is $s_{21}+s_{27}$ in the sum. Exchanging $5,6$ in
\eqref{E2-56} we get another relation.

The second type is the fundamental BCJ relation for each element
$5,6$ and each possible relative ordering of $\{2,3,4\}$ with
remaining elements. For example, let us consider the fundamental BCJ
relation for element $5$. For a given ordering $\sigma'$ in
$P(O\{2,3,4\}\bigcup O\{6\})$, we have a fundamental BCJ relation
\bea \Sl_{\sigma\in P(O\{\sigma'\}\bigcup O\{5\})}\WH
S_5(\sigma)A(\WH1,\sigma,\WH7,...,n)+\Sl_{\gamma\in P(O\{5\}\bigcup
O\{8,...,n-1\})}S_5(\gamma)A(\WH1,\sigma',\WH7,\gamma,n)=0 \eea
We should consider also fundamental BCJ relations for ordering
$\sigma'$ in $P(O\{2,3\}\bigcup O\{6\})$ and a given $\gamma'$ in
$P(O\{4\}\bigcup O\{8,...,n-1\})$, etc. Listing all them together we
have
{\small
\bean 0 &= &\Sl_{\sigma\in P(O\{\sigma'\}\bigcup O\{5\})}\WH
S_5(\sigma)A(\WH1,\sigma,\WH7,...,n)+\Sl_{\gamma\in P(O\{5\}\bigcup
O\{8,...,n-1\})}S_5(\gamma)A(\WH1,\sigma',\WH7,\gamma,n),~~~\sigma'=P(O\{2,3,4\}\bigcup
O\{6\})\nn
0 & = & \Sl_{\sigma\in P(O\{\sigma'\}\bigcup O\{6\})}\WH
S_6(\sigma)A(\WH1,\sigma,\WH7,...,n)+\Sl_{\gamma\in P(O\{6\}\bigcup
O\{8,...,n-1\})}S_6(\gamma)A(\WH1,\sigma',\WH7,\gamma,n),~~~\sigma'=P(O\{2,3,4\}\bigcup
O\{5\})\nn
0 & = & \Sl_{\sigma\in P(O\{\sigma'\}\bigcup O\{5\})}\WH
S_5(\sigma)A(\WH1,\sigma,\WH7,\gamma',n)+\Sl_{\gamma\in
P(O\{5\}\bigcup
O\{\gamma'\})}S_5(\gamma)A(\WH1,\sigma',\WH7,\gamma,n),~~~\left\{\begin{array}{l}
\sigma'=P(O\{2,3,\}\bigcup O\{6\}) \\
\gamma'=P(O\{4\}\bigcup O\{8,...,n-1\})\end{array} \right.\nn
0 & = & \Sl_{\sigma\in P(O\{\sigma'\}\bigcup O\{6\})}\WH
S_6(\sigma)A(\WH1,\sigma,\WH7,\gamma',n)+\Sl_{\gamma\in
P(O\{6\}\bigcup
O\{\gamma'\})}S_6(\gamma)A(\WH1,\sigma',\WH7,\gamma,n),~~~\left\{\begin{array}{l}
\sigma'=P(O\{2,3,\}\bigcup O\{5\}) \\
\gamma'=P(O\{4\}\bigcup O\{8,...,n-1\})\end{array} \right.\nn
0 & = & \Sl_{\sigma\in P(O\{\sigma'\}\bigcup O\{5\})}\WH
S_5(\sigma)A(\WH1,\sigma,\WH7,\gamma',n)+\Sl_{\gamma\in
P(O\{5\}\bigcup
O\{\gamma'\})}S_5(\gamma)A(\WH1,\sigma',\WH7,\gamma,n),~~~\left\{\begin{array}{l}
\sigma'=P(O\{2,\}\bigcup O\{6\}) \\
\gamma'=P(O\{3,4\}\bigcup O\{8,...,n-1\})\end{array} \right.\nn
0 & = & \Sl_{\sigma\in P(O\{\sigma'\}\bigcup O\{6\})}\WH
S_6(\sigma)A(\WH1,\sigma,\WH7,\gamma',n)+\Sl_{\gamma\in
P(O\{6\}\bigcup
O\{\gamma'\})}S_6(\gamma)A(\WH1,\sigma',\WH7,\gamma,n),~~~\left\{\begin{array}{l}
\sigma'=P(O\{2\}\bigcup O\{5\}) \\
\gamma'=P(O\{3,4\}\bigcup O\{8,...,n-1\})\end{array} \right.\nn
\eean
}

Summing all relations from these two types, we get
\bea T&=&\Sl_{\sigma\in P(O\{2,3,4\}\bigcup
\{5,6\})}A(\WH1,\sigma,\WH7,...,n)\nn
&=&-\frac{1}{s_{\WH123456}}\Biggl[\Sl_{\gamma\in P(O\{4\}\bigcup
O\{8,...,n-1\})}(s_{\WH12356}+S_4(\gamma))\Sl_{\sigma\in
P(O\{2,3\}\bigcup \{5,6\})}A(\WH1,\sigma,\WH7,\gamma,n)\nn
&&+\Sl_{\gamma\in P(O\{3,4\}\bigcup
O\{8,...,n-1\})}(s_{\WH1256}+S_3(\gamma)+S_4(\gamma))\Sl_{\sigma\in
P(O\{2\}\bigcup \{5,6\})}A(\WH1,\sigma,\WH7,\gamma,n)\nn
&&+\Sl_{\gamma\in P(O\{2,3,4\}\bigcup O\{8,...,n-1\})}
\left(S_2(\gamma)+S_3(\gamma)+S_4(\gamma)\right)\Sl_{\sigma\in
P( \{5,6\})}A(\WH1,\sigma,\WH7,\gamma,n)\nn
&&+\left\{\Sl_{\gamma\in P(O\{5\}\bigcup
O\{8,...,n-1\})}S_5(\gamma)\Sl_{\sigma\in
P(O\{2,3,4\}\bigcup\{6\})}A(\WH1,\sigma,\WH7,\gamma,n)+\{
5\leftrightarrow 6\}\right\}\nn
&&+\left\{\Sl_{\gamma\in P(O\{5\}\bigcup O\{4\} \bigcup
O\{8,...,n-1\})}S_5(\gamma)\Sl_{\sigma\in
P(O\{2,3\}\bigcup\{6\})}A(\WH1,\sigma,\WH7,\gamma,n)+\{
5\leftrightarrow 6\}\right\}\nn
&&+\left\{\Sl_{\gamma\in P(O\{5\}\bigcup O\{3,4\} \bigcup
O\{8,...,n-1\})}S_5(\gamma)\Sl_{\sigma\in
P(O\{2\}\bigcup\{6\})}A(\WH1,\sigma,\WH7,\gamma,n)+\{
5\leftrightarrow 6\}\right\}\Biggl]~~~\label{E2-T} \eea
Now we can see the large $z$-behavior of $T$. For the first two
lines at the right-handed side of \eqref{E2-T}, although $S$-factor
is ${z\over z}$, partial-ordered permutation sum of $\Sl_{\sigma\in
P(O\{2,3\}\bigcup \{5,6\})}A(\WH1,\sigma,\WH7,\gamma,n)$ and
$\Sl_{\sigma\in P(O\{2\}\bigcup
\{5,6\})}A(\WH1,\sigma,\WH7,\gamma,n)$ do give right behavior by
induction. The third line is the pure permutation sum, but thanking
the pre-factor ${1\over s_{\WH 123456}}$, we get the right result.
For last three lines, although there is only one element $5$ or $6$
in the partial-ordered permutation sum, the pre-factor ${1\over
s_{\WH 123456}}$ provides the wanted reduction of power of $z$.
Overall we have
 \bea T\equiv \Sl_{\sigma\in
P(O\{2,3,4\}\bigcup\{5,6\})}A(\WH1,\sigma,\WH 7,...,n)\sim
\frac{1}{z}\Sl_{\sigma\in P(\{5,6\})}A(\WH1,\sigma,\WH 7,...,n).
\eea
%

\subsection{General proof}
Now let us consider the general case of \eqref{ordered-perm}. As did
in previous example, we need to write down following two types of
relations: the first type is just a single generalized BCJ relation
\eqref{gen-BCJ} with $\{\beta\}=\{2,...,l\}$. For an arbitrary
ordering of the elements in $\{l+1,l+2,...,i-1\}$, we get a BCJ
relation of this type. The second type includes all fundamental BCJ
relations for each element in the set $\{l+1,l+2,...,i-1\}$. For an
arbitrary ordering of the element in
$O\{2,...,g\}\bigcup\{l+1,...,h-1,h+1,...,i-1\}\bigcup
O\{i+1,...,n\}$ with $2\leq g\leq l$\footnote{In the ordering, all
$\{l+1,...,h-1,h+1,...,i-1\}$ are at the left-handed side while
$\{i+1,...,n\}$ are at the right-handed side.}, we get a BCJ
relation of this type
 with $\{\beta\}=\{h\}$.  Summing all the possible BCJ relations of the two
types, we will arrive
\bea
\Sl_{\sigma\in
P(O\{2,...,l\}\bigcup\{l+1,...,i-1\})}A(\WH1,\sigma,\WH i,...,n)=
-\frac{1}{s_{\WH1,2,...,i-1}}\left(I_1+I_2+I_3\right),
\eea
Where $I_1$, $I_2$, $I_3$ are defined as
\bean I_1=\Sl_{g=2}^{l-1}\Sl_{\gamma\in P(O\{g+1,...,l\}\bigcup
O\{i+1,...,n-1\})}\left(s_{\WH12...g(l+1)...(i-1)}+\sum_{i=g+1}^l
S_i(\gamma)\right)\Sl_{\sigma\in
P(O\{2,...,g\}\bigcup\{l+1,...,i-1\})} A(\WH1,\sigma,\WH
i,\gamma,n), \eean
 \bean
 I_2&=&\Sl_{\gamma\in P(O\{2,...,l\}\bigcup O\{i+1,...,n-1\})}\left(S_{2}(\gamma)+...+S_{l}(\gamma)\right)\Sl_{\sigma\in P(\{l+1,...,i-1\})}
A(\WH1,\sigma,\WH i,\gamma,n),
 \eean
 \bean
 I_3=\Sl_{h=l+1}^{i-1}\Sl_{g=2}^{l}\Sl_{\gamma\in P(O\{h\}\bigcup O\{g+1,...,l\}\bigcup O\{i+1,...,n-1\})}S_h(\gamma)\Sl_{\sigma\in P(O\{2,...,g\}\bigcup\{l+1,...,h-1,h+1,...,i-1\})}
A(\WH1,\sigma,\WH i,\gamma,n).
 \eean
  Here $I_1$ comes from both the first and the second types of
  BCJ relations. $I_2$ can only come from
the first type of BCJ relations,while  $I_3$ can only come from the
second type of BCJ relation. All the $I_1$, $I_2$ and $I_3$ are
written in terms of partial-ordered permutation sum, so we  can use
induction. For $I_1$ part, although
 $-\frac{s_{\WH12...gl+1...i-1}}{s_{\WH1,2,...,i-1}}\sim z^0$, the partial-ordered
 permutation sum gives
 \bea I_1\sim \Sl_{\sigma\in P(O\{2,...,g\}\bigcup\{l+1,...,i-1\})}
A(\WH1,\sigma,\WH i,\gamma,n)\sim \frac{1}{z}\Sl_{\sigma\in
P(\{l+1,...,h-1,h+1,...,i-1\})} A(\WH1,\sigma,\WH i,\gamma,n)~.\eea
For the $I_2$ part, because the factor ${1\over s_{\WH 1...(i-1)}}$,
we have immediately
 \bea I_2\sim  \frac{1}{z}\Sl_{\sigma\in P(\{l+1,...,i-1\})}
A(\WH1,\sigma,\WH i,\gamma,n). \eea
For the $I_3$ part, although the number of elements in the
partial-ordered permutation sum is reduced by one, the factor
${1\over s_{\WH 1...(i-1)}}$ provides the wanted another ${1\over
z}$ reduction, thus we have
\bea I_3\sim\frac{1}{z^2}\Sl_{\sigma\in
P(\{l+1,...,h-1,h+1,...,i-1\})} A(\WH1,\sigma,\WH i,\gamma,n)\sim
\frac{1}{z}\Sl_{\sigma\in P(\{l+1,...,h-1,h,h+1,...,i-1\})}
A(\WH1,\sigma,\WH i,\gamma,n). \eea
Summing all contributions from $I_1$, $I_2$ and $I_3$, we proved
\eqref{ordered-perm}.

\section{The combination of cyclic and permutation sums}
In this section, we will prove
\bea \Sl_{\sigma\in Z(\{2,...,i-1\})}\Sl_{\rho\in
P(\{i+1,...,n\})}A(\WH1,\sigma,\WH i,\rho)
\sim\frac{1}{z}\Sl_{\rho\in P(\{i+1,...,n\})}A(\WH1,...,\WH
i,\rho).\label{cyclic-perm} \eea
First let us consider following cyclic sum   for a given ordering of
$\rho$
 \bea &&\Sl_{\sigma\in Z(\{2,...,i-1\})}A(\WH1,\sigma,\WH
i,\rho)=A(\WH1,2,...,i-1,\WH i,\rho)+\Sl_{2\leq k <
i-1}A(\WH1,k+1,...,i-1,2,...,k,\WH i,\rho)\nn
&=&A(\WH1,2,...,i-1,\WH i,\rho)+\Sl_{2\leq k < i-1}\Sl_{\sigma'\in
O^T(2,...,k)\bigcup O\{k+1,...,i-2\}}
(-1)^{k-1}A(\WH1,\sigma',i-1,\WH i,\rho)\nn
&+&\Sl_{2\leq k<i-1}(-1)^{k-1}\Sl_{2\leq l<k}\Sl_{\sigma'\in
P(O\{k+1,...,i-2\}\bigcup O^T\{2,...,l\})} \Sl_{\rho'\in
P(O^T\{l+1,...,k\}\bigcup O\{\rho\})}A(\WH1,\sigma',i-1,\WH
i,\rho'). ~~\label{cy}\eea
where at the second and third lines we have used the KK-relation to
expand $A( \WH i, \rho,\WH 1, k+1,...,i-2, i-1, 2,...,k)$ with
$(i-1)$ and $i$ as pivots. In the expansion, there are two cases.
The first case is that the set $\rho$ and set $\{2,...,i-1\}$ are
separated by $1,i$, while all remaining belong to second case. The
reason we make above separation is that all terms at the second line
sum to zero by exactly same reason as we show the leading part
contribution is zero in the section 3.2.

Having simplified the cyclic sum in \eqref{cy},  the sum combining
cyclic and permutation can be written as
\bea &&\Sl_{\sigma\in Z(\{2,...,i-1\})}\Sl_{\rho\in
P(\{i+1,...,n\})}A(\WH1,\sigma,\WH i,\rho)\nn &=&\Sl_{2\leq
k<i-1}(-1)^{k-1}\Sl_{2\leq l<k}\Sl_{\sigma'\in
P(O\{k+1,...,i-2\}\bigcup O^T\{2,...,l\})} \Sl_{\rho'\in
P(O^T\{l+1,...,k\}\bigcup\{i+1,...,n\})}A(\WH1,\sigma',i-1,\WH
i,\rho'). \eea For given $k$, $l$ and $\sigma'$, the sum
$\Sl_{\rho'\in
P(O^T\{l+1,...,k\}\bigcup\{i+1,...,n\})}A(\WH1,\sigma',i-1,\WH
i,\rho')$ is nothing but the partial-ordered permutation sum
discussed in previous section. As we have shown in the previous
section, the $z$-dependence should be suppressed by factor
$\frac{1}{z}$. Thus the behavior \eqref{cyclic-perm} is proved.

\section{Conclusion}
In this note, using KK-relations and generalized BCJ relations, we
have proved the large-$z$ behavior of cyclic sum and combination sum
under BCFW recursion, which were first analyzed in
\cite{Boels:2011tp,Boels:2011mn} through explicit Feynman diagram analysis. Our
proof shows that these better convergent behaviors are natural
consequences of the familiar boundary behavior of single pair
BCFW-deformation, in the sense that
they do not imply new nontrivial relations
between amplitudes. However, as pointed by Boels and
Isermann\cite{Boels:2011tp,Boels:2011mn}, these behaviors do have
important applications in understanding various properties of
amplitudes at one-loop level.

A new result presented in our paper is the boundary behavior of
partial-ordered permutation sum, which played a crucial role in our
proof of the large-$z$ behavior
of the combination sum. It would be interesting to see if the
partial-ordered sum imposes further constraints on loop-level
amplitudes similar to the vanishing conditions on box, triangle and bubble
coefficients at one-loop imposed by cyclic and permutation sums.

We note that following the same
reasoning in theories such as  $\mathcal{N}=4$ SYM where
KK and BCJ relations hold amplitudes are expected to present similar convergent
behavior. However theories where gauge field couples to
 gravity require special attentions. In these cases despite KK relations
remain valid it was argued\cite{Chen:2010ct}
that BCJ relations cannot assume the usual form as in pure gauge theory.
Having only KK relations at our disposal,
we can derive
  the large-$z$ behavior of the cyclic sum in the adjacent case,
  while other types of sums in this theory await further studies.

\subsection*{Acknowledgements}
Y. J. Du is supported in part by the NSF of China Grant No.11105118,
No.11075138. B. Feng is supported by fund from Qiu-Shi, the
Fundamental Research Funds for the Central Universities with
contract number 2010QNA3015, as well as Chinese NSF funding under
contract No.10875104, No.11031005,  No.11135006, No. 11125523.


\end{document}